\documentclass[twocolumn]{aastex631}  
\usepackage[utf8]{inputenc}
\usepackage{xcolor}
\usepackage{amsmath}
\usepackage{multirow}
\usepackage{xcolor}

\newcommand{\del}[1]{\textcolor{magenta}{{\iffalse{#1}\fi}}}

\shorttitle{Rapid Orbital Decay of WX Cen: a Surrounding Circumbinary Disk}
\shortauthors{Liu, Liu \& Chen}

\begin{document}


\title{Rapid Orbital Decay of Supersoft X-Ray Source WX Cen: a Surrounding Circumbinary Disk}

\author[0009-0005-7325-2756]{Zhi-Qiang Liu}
\affil{School of Science, Qingdao University of Technology, Qingdao 266525, People’s Republic of China;chenwc@pku.edu.cn}
\author[0000-0002-7502-4352]{Wei-Min Liu}
\affil{School of Physics and Electrical Information, Shangqiu Normal University, Shangqiu 476000, People’s Republic of China}
\author[0000-0002-0785-5349]{Wen-Cong Chen}
\affil{School of Science, Qingdao University of Technology, Qingdao 266525, People’s Republic of China;chenwc@pku.edu.cn}
\affil{School of Physics and Electrical Information, Shangqiu Normal University, Shangqiu 476000, People’s Republic of China}



\begin{abstract}
WX Cen is most likely a candidate of compact binary supersoft X-ray sources, which consists of a white dwarf (WD) and a donor star that fills the Roche lobe. Recently, this source was detected to be experiencing a rapid orbital decay at a rate of $\dot{P} =-(4.4\pm0.4)\times10^{-7}~ \rm days~\rm{yr}^{-1}$. According to the mass function and optical eclipses, the donor-star mass can be constrained to be $0.41-0.44$,  $0.47-0.50$, and $0.55-0.59~M_\odot$ when the WD mass is $0.7$, $0.9$, and $1.2~M_\odot$, respectively. The observed orbital period derivative cannot be produced by the angular momentum loss due to mass loss during the accretion of the WD, magnetic braking (MB) mechanisms, including standard MB, convection and rotation-enhanced MB, and anomalous MB prescriptions. We propose that the rapid orbital decay of WX Cen is caused by the tidal torque that originates from the resonant interaction between the binary and a surrounding circumbinary (CB) disk. Detailed stellar evolution models indicate that a WD binary with a $2.5 \times 10^{-7}~M_{\odot}$ CB disk can evolve toward a WX Cen-like system, which has an orbital period derivative of $\dot{P}=-4.0\times10^{-7}~ \rm days~\rm{yr}^{-1}$ and a relatively high mass-transfer rate of $5.3\times10^{-7}~M_\odot\rm yr^{-1}$ that can trigger a stable hydrogen burning process on the surface of the WD. 
\end{abstract}


\section{Introduction}
\label{sec:intro} 
Supersoft X-ray sources (SSSs) are a class of soft X-ray objects with a blackbody temperature range of $20-100$ eV and a luminosity larger than $10^{35}~\rm erg~s^{-1}$ \citep{grei96}. Einstein Laboratory discovered the first SSS in the Large Magellanic Cloud \citep{long81}. It is now generally thought that SSSs are binary systems including massive white dwarfs (WDs) which steadily burn nuclear fuel on their surface due to accretion from a main sequence (MS) or subgiant donor star at a rate near or above the Eddington accretion rate \citep{heuv1992,kaha1997}. High mass accretion rates could eventually lead to massive WDs growing beyond the Chandrasekhar limit and undergoing an explosion or collapse. Therefore, SSSs are proposed to be the long-sought progenitors of Type Ia supernovae \citep{li97}.

The variable star WX Cen was initially proposed to be a possible optical counterpart of the hard X-ray transient source Cen XR-2 \citep{egge68}. Subsequently, WX Cen was identified as a Wolf–Rayet star of type WN 7 according to its spectral characteristics \citep{huch81}. By time-resolved spectro-photometric observations, this source was identified as a binary system with an orbital period of 10.0 hours \citep{daiz1995}. Based on spectroscopic observations, \citet{oliv04} refined the orbital period of WX Cen to $P = 0.4169615(\pm 22)~\rm{days}$. The spectroscopic and photometric characteristics indicated that WX Cen is most likely one of the Galactic counterparts of compact binary SSSs \citep{patt98,oliv04}.

By using four newly determined eclipse timings together with archival data, \citet{Qian2013} found that the orbital period of WX Cen is decreasing at a rate 
 $\dot{P} =-5.15\times10^{-7}~ \rm days~\rm{yr}^{-1}$. Recently, the orbital period derivative was refined to $\dot{P} =-(4.4\pm0.4)\times10^{-7}~ \rm days~\rm{yr}^{-1}$ using times of minima determined from the light curves of the Transiting Exoplanet Survey Satellite and the American Association of Variable Star Observers and archival data \citep{zang23}. When the mass is transferred from the massive donor star to the less massive WD, the orbital period decreases. However, the WD mass was inferred as $M_{\rm wd}\sim 0.9~M_\odot$ if the intrinsic velocity ($V=6100~ \rm km~s^{-1}$) of the jet, derived from the satellite-like events at the Balmer lines, originates from the escape velocity of the WD \citep{oliv04}. Using the WD radial-velocity semi-amplitude $K_{\rm wd} = 108\pm3~\rm km~s^{-1}$ \citep{daiz1995,oliv04} and $M_{\rm wd}\sim 0.9~M_\odot$, the donor-star mass was derived as $M_{\rm d}\sim0.6~M_\odot$ from the mass function equation \citep{zang23}. In principle, the orbit should expand when the mass is transferred from the less massive donor star to the more massive WD. Therefore, WX Cen should experience an efficient angular momentum loss (AML) during its evolution.

The orbital decay of WX Cen was thought to originate from the AML via magnetic wind from the donor star and/or from the accretion disk \citep{zang23}. To account for the observed period derivative ($\dot{P} =-(4.4\pm0.4)\times10^{-7}~ \rm days~\rm{yr}^{-1}$) of WX Cen, the total wind-loss rate ($\dot{M}_{\rm w}$) and the Alfven radius ($R_{\rm A}$) have to satisfy $\dot{M}_{\rm w}=-5.0\times10^{-9}(20~R_\odot/R_{\rm A})^2~M_\odot\rm yr^{-1}$ \citep{zang23}. Even if WX Cen possesses a large Alfven radius ($R_{\rm A}=20~R_\odot$) similar to the Sun, it still requires an extremely high wind-loss rate of $\dot{M}_{\rm w}=-5.0\times10^{-9}~M_\odot\rm yr^{-1}$. However, the stellar wind-loss rate with a tidal enhancement factor is \citep{tout88}
\begin{equation}
\begin{aligned}
\dot{M}_{\rm w}=-4.0\times10^{-13}~M_\odot{\rm yr}^{-1}\frac{r_{\rm d}l_{\rm d}}{m_{\rm d}}\times \\
\left(1+10^4\times {\rm min}\left[\left(\frac{r_{\rm d}}{r_{\rm L}}\right)^6, \frac{1}{2^6}\right]\right),
\end{aligned}
\end{equation}
where $m_{\rm d}$, $r_{\rm d}$, $l_{\rm d}$, and $r_{\rm L}$ are the mass, the radius, the luminosity, and the Roche-lobe radius of the donor star in solar units, respectively. This enhanced wind-loss rate is an extreme and controversial model because of the lack of observational evidence and theoretical support. Using the same parameters as \cite{zang23}, the wind-loss rate is $\dot{M}_{\rm w}=-8.5\times10^{-11}~M_\odot\rm yr^{-1}$, which is approximately two orders of magnitude smaller than the one derived by \cite{zang23}. Therefore, it seems that magnetic winds are unlikely to be responsible for the orbital decay observed in WX Cen.

Furthermore, the luminosity of SSSs was thought to arise from a stable nuclear burning of accreting hydrogen on the surface of WDs \citep{heuv1992}. To achieve a stable hydrogen burning process in WX Cen, an extremely high accretion rate of $\sim10^{-7}~M_\odot\rm yr^{-1}$ is required. For a low-mass donor star with a mass of $\sim0.6~M_\odot$, it is incredible to produce such a high mass-transfer rate unless an efficient AML mechanism exists \citep{liu19}. Therefore, the formation and evolution of WX Cen remain mysterious.

This work investigates the formation of the SSS candidate WX Cen and attempts to account for its rapid orbital decay and the high mass accretion rate. We firstly constrain two components' masses of WX Cen in Section 2. In Section \ref{sec:Analyze}, we analyze the orbital evolution of WX Cen. Section \ref{sec:three} diagnoses the contributions of three different magnetic braking (MB) prescriptions on the period derivative. Section \ref{sec:CB disk} studies the possibility of a surrounding CB disk and presents a detailed stellar evolution model for the formation of WX Cen. In Sections \ref{sec:discussion} and \ref{sec:summary}, we give a brief discussion and summary, respectively.

\section{Constraining the donor-star mass of WX Cen}
\label{sec:uncontain}
Actually, the estimated WD mass ($\sim0.9~M_\odot$) of WX Cen was based on the following assumptions: the Balmer lines come from ejected gas; the orbital inclination $i=55^\circ$; the intrinsic 'jet' velocity
represents the escape velocity of a WD \citep{oliv04}. In particular, the orbital inclination of WX Cen was adopted as the same value as that of QR And because of the similar observed light-curve amplitude, light-curve shape, and photometric behavior \citep{oliv04}. \cite{zang23} derived the donor-star mass of $\sim0.6~M_\odot$ according to the WD mass of $\sim0.9~M_\odot$, the mass function, and an assumed orbital inclination $i=55^\circ$. Therefore, the masses of the two components are highly uncertain due to the extreme uncertainties of these assumptions.

The radial velocity semi-amplitude of the WD was measured to be $K_{\rm wd} = 108~\rm km~s^{-1}$ \citep{daiz1995,oliv04}. Therefore, the mass function of WX Cen can be derived as 
\begin{equation}
f(M_{\rm wd},M_{\rm d},i) = \frac{M_{\rm d}^3 \sin^3 i}{(M_{\rm wd} + M_{\rm d})^2} = 0.054~M_\odot,
\end{equation}
where $M_{\rm wd}$ and $M_{\rm d}$ are the WD mass and donor-star mass, respectively. To produce optical eclipses, the orbital inclination should satisfy
\begin{equation}
{\rm sin}i =\sqrt{1-{\rm cos}^2i}>\sqrt{1-\left(\frac{R_{\rm d}}{a}\right)^2 },
\end{equation}
where $R_{\rm d}$ and $a$ are the donor-star radius and orbital separation, respectively. Since the donor star fills the Roche lobe, we use the Roche-lobe radius instead of the donor-star radius.

Taking three typical WD masses, we plot the function relation between ${\rm sin}i$ and $M_{\rm d}$ of WX Cen in Figure 1 according to equation (2). Subsequently, 
equation (3) can constrain the donor-star mass to be $0.41-0.44$,  $0.47-0.50$, and $0.55-0.59~M_\odot$ when $M_{\rm wd}=0.7$, $0.9$, and $1.2~M_\odot$ (see also Table 1), respectively. To produce optical eclipses, ${\rm sin}i>0.94$ and the orbital inclination $i>70^\circ$, which is larger than the one ($i=55^\circ$) adopted by \cite{oliv04} and \cite{zang23}. In Case B (with a WD mass of $0.9~M_\odot$), our constraint is at least $0.1~M_\odot$ lower than the  donor-star mass ($\sim0.6~M_\odot$) estimated by \cite{zang23}. Furthermore, the thermal timescale mass-transfer from the more massive donor star to the less massive WD can be fully ruled out, although it can produce a negative $\dot{P}$.

\section{Analysis on the orbital evolution of WX Cen}
\label{sec:Analyze} 
The orbital angular momentum of a WD binary is $J=M_{ \rm wd}M_{\rm d}/(M_{\rm wd}+M_{\rm d}) a^{2}2\pi/P$, where $P$ is the orbital period.  During the mass transfer, the mass-loss rate of the donor star is $\dot{M}_{\rm d} = \dot{M}_{\rm w}-\dot{M}_{\rm tr}$, where $\dot{M}_{\rm w}$ and $\dot{M}_{\rm tr}$ ($>0$) are the stellar wind-loss rate and mass-transfer rate, respectively. Considering the accretion efficiency of the WD is $f$, the accretion rate of the WD is $\dot{M}_{\rm wd}=-f\dot{M}_{\rm d}$. Differentiating the above equation and combining with Kepler’s third law ($G(M_{\rm wd}+M_{\rm d})/a^{3}=4\pi^{2}/P^{2}$, where $G$ is the gravitational constant), the orbital-period derivative of WX Cen satisfies
\begin{equation}
\frac{\dot{P}}{P} = 3\frac{\dot{J}}{J} - 3\frac{\dot{M}_{\rm d}}{M_{\rm d}}\left[ 1 - qf - \frac{q(1 - f)}{3(1 + q)} \right],
\label{pdot}
\end{equation}
where $q=M_{\rm d}/M_{\rm wd}$ and $\dot{J}$ are the mass ratio and the rate of orbital AML of the binary, respectively. 

The first term on the right-hand side of Equation (\ref{pdot}) causes an orbital decay effect since $\dot{J}<0$, but the second term produces a positive $\dot{P}$ if $g(q,f) = 1 - qf - \frac{q(1 - f)}{3(1 + q)}>0$. Therefore, it strongly depends on the competition between these two effects whether the orbit of WX Cen is decaying as observed recently. We first diagnose whether different AML mechanisms are sufficient to cause the observed orbital decay rate. Thus, we calculate the orbital period derivative using the following equation 
\begin{equation}
\label{equ:3}
\dot{P} = \frac{3\dot{J}P}{J}.
\end{equation}

In general, the rate of AML of a binary system is $\dot{J}=\dot{J}_{\rm gr}+\dot{J}_{\rm mb}+\dot{J}_{\rm ml}$, where $\dot{J}_{\rm gr}, \dot{J}_{\rm mb}$, and $\dot{J}_{\rm ml}$ are the rates of AML caused by gravitational radiation (GR), magnetic braking (MB), and mass loss , respectively. In this Section, we firstly evaluate the contribution of GR and mass loss. The period derivative due to GR is given by 
\begin{equation}
\label{equ:4}
\dot{P}_{\rm gr} = -\frac{96 G^3}{5 c^5} \frac{M_{\rm wd} M_{\rm d} (M_{\rm wd} + M_{\rm d})}{a^4} P,
\end{equation}
where $c$ is the speed of light in vacuum. According to our constrained donor-star masses, the period derivative of WX Cen due to GR is $\dot{P}_{\rm gr}=-(0.9-2.0)\times10^{-11}~\rm days\ yr^{-1}$ (see also Table 1), which is four orders of magnitude smaller than the observed value.

During the accretion of WDs, strong hydrogen and helium shell flashes can cause mass loss. The rate of AML due to mass loss can be expressed as
\begin{equation}
\dot{J}_{\rm ml} = \frac{\alpha + \beta q^2 + \delta\gamma (1+q)^2}{1+q} \frac{\dot{M}_{\rm d}}{M_{\rm d}} J,
\label{equ:5}
\end{equation}
where $\alpha$, $\beta$, and $\delta$ represent the fractions of mass loss from the donor star in the form of stellar winds, the mass ejected from the vicinity of the WD, and from a surrounding circumbinary disk with a radius of $a_{\rm r} = \gamma^2 a$ \citep{taur23}. Taking $\alpha=\delta=0$, the orbital-period derivative can be derived as
\begin{eqnarray}
\dot{P}_{\rm ml}=-3.75\times 10^{-8} \left( \frac{\dot{M}_{\rm d}}{ -10^{-7}~M_\odot \, {\rm yr}^{-1}} \right)\left( \frac{\beta}{0.9} \right) \nonumber \\
\left( \frac{q}{0.5} \right) \left(\frac{1.5~M_\odot}{M_{\rm wd}+M_{\rm d}}\right)\left( \frac{P}{0.417~\rm days} \right) \, \rm days \, yr^{-1}.
\label{equ:6}
\end{eqnarray}
Setting $\beta=0.9$ and $\dot{M}_{\rm d}=-10^{-7}~M_\odot \, \rm yr^{-1}$ (which is an extremely high mass loss rate unless there is an efficient AML mechanism), the orbital-period derivative of WX Cen due to the mass loss is estimated to be $\dot{P}_{\rm ml}=-(2.9-6.2) \times 10^{-8} ~\rm days~ yr^{-1}$ (see also Table 1), which is an order of magnitude smaller than observed $\dot{P}$.

\begin{table*}
\centering
\caption{Constrained Donor-star Masses and Calculated Orbital Period Derivatives for Different Physical Processes}
\label{tab:binary_para}
\tabletypesize{\small} % 
\begin{tabular}{l c c c c c c c c c}
\hline\hline % 
 \colhead{Cases} & \colhead{$M_{\rm wd}$} & \colhead{$M_{\rm d}$} & \colhead{$-\dot{P}_{\rm gr}$} & \colhead{$-\dot{P}_{\rm ml}$} & \colhead{$-\dot{P}_{\rm smb}$} & \colhead{$-\dot{P}_{\rm carb}$} & \colhead{$-\dot{P}_{\rm amb}$}\\
& $(M_\odot)$ & $(M_\odot)$  & $(10^{-11}\,\mathrm{days\,yr}^{-1})$ & $(10^{-8}\,\mathrm{days\,yr}^{-1})$ & $(10^{-9}\,\mathrm{days\,yr}^{-1})$ & $(10^{-9}\,\mathrm{days\,yr}^{-1})$ & $(10^{-8}\,\mathrm{days\,yr}^{-1})$  \\
\hline
 A & 0.7 & $0.41-0.44$ &  $0.9-1.0$ & $5.9-6.2$ & $2.6-2.9$    & $12.1-12.4$ & $6.3-6.4$ \\
 B & 0.9 & $0.47-0.50$ &  $1.3-1.4$ & $4.3-4.5$ & $2.5-2.8$    & $10.1-10.3$ & $5.1-5.2$ \\
 C & 1.2 & $0.55-0.59$ &  $1.9-2.0$ & $2.9-3.1$ & $2.5-2.8$    & $8.3-8.4$   & $4.0-4.1$ \\
\hline % 
\end{tabular}
\label{tab:1}
\end{table*}

\section{Three different MB prescriptions}
\label{sec:three}
\subsection{Standard MB prescription}
In binary systems with low-mass donor stars, MB is an important mechanism for extracting orbital AML. The coupling between the stellar wind and a magnetic field can cause low mass stars to spin down. Assuming that the donor star being braked is tidally synchronized, the standard MB prescription proposed a rate of AML as \citep{Rappaport1983}
\begin{equation}
\dot{J}_{\rm{smb}}=-6.82\times10^{34}\left(\frac{M_{\rm{d}}}{M_{\odot}}\right)\left(\frac{R_{\rm{d}}}
{R_{\odot}}\right)^{\gamma_{\rm mb}}\left(\frac{P}{1~\rm d}\right)^{-3}\rm~g~cm^2s^{-2}, 
\label{equ:7}
\end{equation}
where $\gamma_{\rm mb}$ is a dimensionless parameter from 0 to 4. Adopting $\gamma_{\rm mb}=4$ \citep{Verbunt1981}, the standard MB prescription can produce an orbital period derivative of $\dot{P}_{\rm smb}= -(2.5-2.9)\times 10^{-9}~\rm days\,yr^{-1}$ for WX Cen (see also Table 1). Therefore, the standard MB prescription cannot be responsible for the observed $\dot{P}$.

\subsection{CARB MB prescription}
To reproduce the observed mass-transfer rates and orbital periods for all observed persistent neutron star (NS) low-mass X-ray binaries (LMXBs) with detected mass ratios, \citet{Van2019} proposed a convection- and rotation-boosted (CARB) MB prescription, which includes the influence of the donor stars' rotation on the stellar wind velocity, the donor stars' convective-turnover timescale, and rotation on their surface magnetic field. In the CARB MB prescription, the rate of AML is given by \citep{Van2019} 
\begin{eqnarray}
\dot{J}_{\rm{carb}} = -\frac{
2}{3} |\dot{M}_{\rm{w}}|^{-1 / 3} R_{\rm{d}}^{14 / 3}\left(v_{\rm{esc}}^{2}+2 \Omega^{2} R_{\rm{d}}^{2} / K_{2}^{2}\right)^{-2 / 3} \nonumber \\
\quad \times \Omega_{\odot} B_{\odot}^{8 / 3}\left(\frac{\Omega}{\Omega_{\odot}}\right)^{11 / 3}\left(\frac{\tau_{\rm{conv}}}{\tau_{\odot, \rm{conv}}}\right)^{8 / 3},
\label{equ:9}
\end{eqnarray}
where $v_{\rm esc}$, $\Omega=2\pi/P$, and $\tau_{\rm conv}$ are the escape velocity, the angular velocity and the turnover time of convective eddies of the donor star, respectively; $K_2=0.07$ is a constant resulted from a grid of simulations \citep{Victor2015}; $\Omega_{\odot}\approx 3\times 10^{-6}~\rm s^{-1}$ and $B_{\odot}=1~\rm G$ represent the rotation rate and magnetic field strength of the solar surface; $\tau_{\rm conv,\odot}=2.8\times10^6~\rm s$ is the convective turnover time of the Sun.

Based on our constrained donor-star masses, the escape velocity of the donor star of WX Cen is calculated to be $v_{\rm esc}=(4.4-5.0)\times10^7~\rm cm~s^{-1}$. Adopting an enhanced stellar wind-loss rate of $\dot{M}_{\rm w}=-8.5\times10^{-11}~M_\odot\rm yr^{-1}$ (which arises from the estimation for a $0.6~M_\odot$ donor star according to equation 1) and $\tau_{\rm{conv}} = 3~\tau_{\odot}$ ($\tau_{\rm{conv}}\approx 1-3~\tau_{\odot}$ for a $0.4-0.6~M_\odot$ star from a detailed stellar evolution model), the rate of AML of WX Cen can be calculated to be $\dot{J}_{\rm carb}=-(0.8-1.2)\times10^{36}~\rm g\,cm^2s^{-2}$. Therefore, the CARB MB prescription predicts an orbital period derivative as $\dot{P}_{\rm carb}\approx -(8.3-12.4)\times 10^{-9}~\rm days\,yr^{-1}$ (see also Table 1), which is $1-2$ orders of magnitude smaller than the observed value of WX Cen.

\subsection{Anomalous MB prescription}
\citet{Justham2006} proposed that black hole (BH) LMXBs evolve from BH intermediate-mass X-ray binaries through an anomalous MB mechanism, which originates from the coupling between the strong magnetic field of Ap/Bp stars and the irradiation-driven wind induced by X-ray radiation. The anomalous MB prescription predicts a rate of AML as
\begin{eqnarray}
\dot{J}_{\rm amb} = -\frac{2\pi B_{\rm s}R_{\rm d}^{13/4} |\dot{M}_{\rm w}|^{1/2}} {(G M_{\rm d})^{1/4}P},
\label{equ:11}
\end{eqnarray}
where $B_{\rm s}$ denotes the surface magnetic field strength of the donor star.

Assuming that WX Cen evolves from a WD binary including an intermediate-mass Ap/Bp star with an extremely strong surface magnetic field of $B_{\rm s}=10^4~\rm G$ and take 
an enhanced stellar wind-loss rate of $\dot{M}_{\rm w}=-8.5\times10^{-11}~M_\odot\rm yr^{-1}$, we can derive $\dot{J}_{\rm amb}= -(4.1-5.5)\times10^{36}~\rm g\,cm^2s^{-2}$ if the decay of magnetic field is ignored. Therefore, the anomalous MB prescription can cause an orbital period derivative of $\dot{P}_{\rm amb}=-(4.0-6.4)\times 10^{-8}~\rm days\,yr^{-1}$ (see also Table 1), which is still an order of magnitude smaller than the observed $\dot{P}$.

In summary, none of the three MB prescriptions adequately explains the rapid orbital decay of WX Cen. An efficient AML mechanism is thus required to account for its unusually large orbital period derivative.

\section{A surrounding CB disk}
\label{sec:CB disk}
\subsection{CB disk model}

There may exist CB disks surrounding binary systems, which could be fed by mass loss during mass transfer \citep{den1973,den1994} or a single outburst or successive outbursts \citep{xu18}, or could be the remnants of the common envelope phase \citep{Spruit2001,Taam2001}. The resonant interaction between the binary and the CB disk can produce a tidal torque, extracting orbital angular momentum from binary systems. The influence of surrounding CB disks on the orbital evolution was extensively investigated in some binary systems such as cataclysmic variables \citep{Spruit2001,Taam2001}, black hole LMXBs \citep{Chen2006,chen2019}, Algol binaries \citep{chen.el2006}, detached binary systems \citep{chen2017}, the eclipsing post-common envelope binary DE CVn \citep{Han2018}, and NS LMXB 2A 1822-371 \citep{wei2023}. 

The rate of AML through a surrounding CB disk can be written as \citep{chen2019}
\begin{eqnarray}
\dot{J}_{\rm cb} = - M_{\rm cb} \alpha \left( \frac{H}{R} \right)^2 \frac{a^3}{R} \Omega^2,
\label{equ:14}
\end{eqnarray}
where $M_{\rm cb}$ is the mass of the CB disk, $\alpha$ is the viscous parameter,  $H$ and $R$ are the thickness and the half angular-momentum radius of the CB disk, respectively. According to equation (3), a surrounding CB disk produces an orbital period derivative as \citep{chen2019}
\begin{equation}
\dot{P}_{\rm cb} = - 6\pi M_{\rm cb} \alpha\left( \frac{H}{R} \right)^2 \frac{a}{R} \frac{1}{\mu},
\label{equ:pdot_cb}
\end{equation}
where $\mu={M_{\rm wd} M_{\rm d}}/({M_{\rm wd} + M_{\rm d}})$ is the reduced mass of the system.

\begin{figure}
\centering
\includegraphics[width=1.0\linewidth,trim={0 0 0 0},clip]{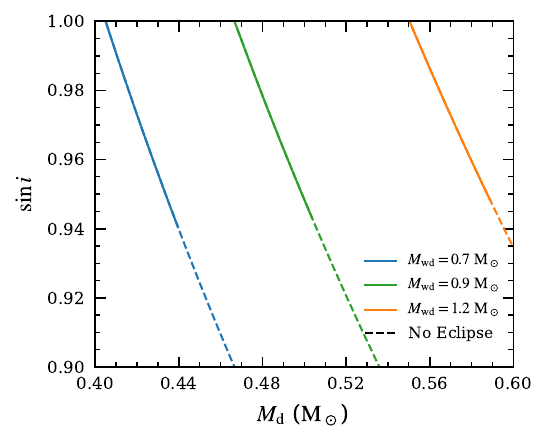}
\caption{Relation between $\sin i$ and $M_{\rm d}$ for WX Cen drived from the WD mass function. The blue, green, and orange curves correspond to WD masses of $0.7$, $0.9$, and $1.2\ M_\odot$, respectively. Solid curves represent the parameter range that can result in optical eclipses.}
\label{fig:1}
\end{figure}

For WX Cen with a circular orbit, we take $r_{\rm in}=1.7a$ and $r_{\rm out}=10a$ \citep{oome20}, hence the half angular-momentum radius of the CB disk is $R=\sqrt{r_{\rm in}r_{\rm out}}\approx4.12a$. Therefore, the orbital period derivative predicted by the CB disk model depends on the degenerate disk parameter $M_{\rm cb} \alpha(H/R)^2$ and the reciprocal of the reduced mass. For simplicity, we set $\alpha=0.1$ and $H/R=0.1$ in the following calculations. Figure \ref{fig:1} plots the predicted $-\dot{P}_{\rm  cb}$ under different reduced masses and CB disk masses. It is clear that a heavier CB disk produces a larger $-\dot{P}_{\rm  cb}$ for the same reduced mass. A CB disk with a mass of $\sim10^{-7}~M_{\odot}$ can account for the orbital period derivative observed in WX Cen. However, it is an open problem whether a surrounding CB disk can produce an extremely high accretion rate of $\sim10^{-7} ~M_\odot \rm \,yr^{-1}$ to achieve a stable hydrogen burning process in WX Cen. Therefore, it is inevitable to perform a detailed binary evolution model for the formation of WX Cen.

\subsection{Stellar Evolution Code}
\label{sec:code}
A binary update version in the Modules for Experiments in Stellar Astrophysics \citep[MESA; version r-12115;][]{Paxton_2011,Paxton_2013,Paxton_2015,Paxton_2018,Paxton_2019} is used to model the formation and evolution of WX Cen, whose progenitor is assumed to be a binary system consisting of a CO WD and a low-mass MS star in a circular orbit. In the calculation, the WD is treated as a point mass, and the code only simulates the nuclear burning in the MS star and the orbital evolution of the system. The initial MS companion star is assumed to have a solar compositions ($X = 0.70,Y = 0.28, Z = 0.02$). For different CB disk masses, the code continuously simulates the evolution of WD-MS binaries with varying initial WD masses, donor-star masses, and orbital periods until the calculated parameters match the observed values.

During the evolution of the MS companion star, we adopt the “Reimers” wind setting options with a scaling factor of 1.0 (which results in a very efficient wind-loss rate) in the schemes including $hot\_wind\_scheme$, $cool\_wind\_AGB\_scheme$, and $cool\_wind\_RGB\_scheme$ \citep{reim75}. Once the donor star fills the Roche lobe,  the hydrogen-rich envelope is transferred onto the WD at a rate $\dot{M}_{\rm tr}$. Therefore, the mass-loss rate of the donor star is $\dot{M}_{\rm d}=-\dot{M}_{\rm tr}+\dot{M}_{\rm w}$. The accreted material is heated and compressed on the surface of the CO WD, eventually triggering thermonuclear reactions. An unstable nuclear burning could result in H and He flashes, in which a part of the accreted material is expelled from the WD surface. As a consequence, the growth of the WD mass depends on the accumulation efficiency with which the accreted hydrogen is converted into helium and the accumulated helium is converted into CO elements. In general, these two accumulation efficiencies are related to the hydrogen abundance in the accreted matter, the WD mass, and the mass-transfer rate \citep{Kato2004,Han2004,Hachisu1999,Kovetz1994}. The estimation of the accumulation efficiencies of hydrogen and helium elements is highly uncertain. Following \citet{Chen2022}, the mass-growth rate of the WD is assumed to be $\dot{M}_{\rm wd} =\eta\dot{M}_{\rm tr}$, and the mass-growth efficiency is taken to be $\eta = 0.1$, which is a time-average efficiency. The excess material is assumed to be ejected from the surface of the WD and form an isotropic fast wind, carrying away the specific orbital angular momentum of the WD. Our inlists are available on Zenodo: \dataset[10.5281/zenodo.20838363]{10.5281/zenodo.20838363}.

Once the mass transfer initiates, a small fraction of the mass loss is assumed to form a surrounding CB disk with a constant mass. Meanwhile, we also consider three AML mechanisms, including gravitational wave radiation, mass loss, and the MB process with a standard prescription ($\gamma_{\rm mb}=4$). The MB mechanism is effective only if donor stars possess both a convective envelope and a radiative core.

\begin{figure}
\centering
\includegraphics[width=1.0\linewidth,trim={0 0 0 0},clip]{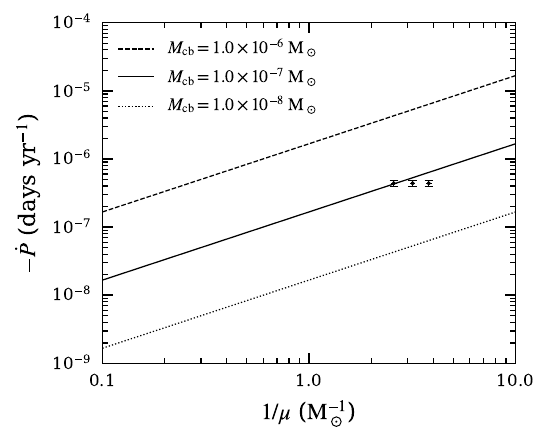}
\caption{Predicted orbital-period derivatives by surrounding CB disks in the $\dot{P}_{\rm cb}$ vs. $1/\mu$ diagram under different CB-disk masses. The solid, dashed, and dotted curves correspond to CB-disk masses of $10^{-7}$, $10^{-6}$, and $10^{-8}~M_\odot$, respectively. Three solid circles with error bars represent the WX Cen-like system with two-components' masses same to Cases A, B, and C.}  
\label{fig:2}
\end{figure}

\subsection{Simulated Results} 
\label{sec:simulated}
To investigate the formation of WX Cen, we perform a grid of binary evolution models for different WD binaries, in which the ranges of initial orbital periods, initial donor-star masses, and initial WD masses are $P_{\rm i} = 0.5-4.0 ~\rm days$, $ M_{\rm d,i} = 1.0-2.0~M_\odot$ and $M_{\rm wd,i} = 0.75-0.85~M_\odot$, respectively. The mass of the CB disk is assumed to be $M_{\rm cb}=10^{-9}-10^{-5}~M_\odot$. If a simulation can reproduce the observed (orbital period and orbital-period derivative) and our constrained parameters (WD and donor-star masses) of WX Cen, we then fine-tune the initial orbital period and initial donor-star mass to obtain the best model. For simplicity, we only perform a detailed stellar evolution model for Case B. For Cases A and C, the initial parameters can be tuned to achieve agreement with the observations and constraint.

Our simulations find that a WD binary with $M_{\rm wd,i} = 0.84 ~M_\odot$, $M_{\rm d,i} = 1.0~M_\odot$, and $P_{\rm i} =3.26~\rm days$ can evolve into a WX Cen-like system when $M_{\rm cb}=2.5\times 10^{-7}~M_\odot$. At the current orbital period of 0.417 days, the orbital period derivative of the WX Cen-like system is $\dot{P}=-4.0\times10^{-7}~\rm days\,yr^{-1}$, which is consistent with the observed one. Meanwhile, our model can approximately reproduce the constrained WD mass and donor-star mass.

\begin{figure}
\centering
\includegraphics[width=1.0\linewidth,trim={0 0 0 0},clip]{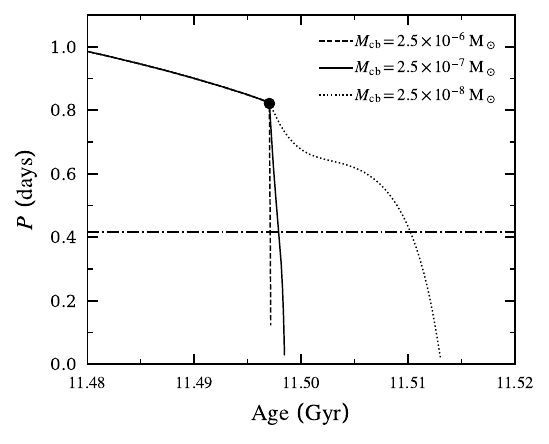}
\caption{Evolution of a WD binary with $M_{\rm wd,i}=0.84~M_\odot$, $M_{\rm d,i}=1.0~M_\odot$, and $P_{\rm i}=3.26~\rm days$ in the orbital period vs. stellar age diagram. The solid, dashed, and dotted curves correspond to CB disk masses of $2.5 \times 10^{-7}$, $2.5 \times 10^{-6}$, and $2.5 \times 10^{-8}~M_\odot$, respectively. The horizontal dashed-dotted line and solid circle represent the current orbital period of WX Cen and the onset of the mass transfer, respectively.}
\label{fig:3}
\end{figure}

Figure \ref{fig:3} shows the evolutionary track of the system mentioned above in the orbital period versus stellar age diagram. In the detached stage, the MB mechanism dominates the orbital evolution until the orbital period decreases to $0.82~\rm days$. At a stellar age of $11.4970~\rm Gyr$, the donor star fills the Roche lobe and begins mass transfer. A heavier CB disk naturally produces a more efficient AML, driving the system to evolve to the current orbital period ($P = 0.417~\rm days$) of WX Cen in a shorter timescale. When $M_{\rm cb}=2.5\times 10^{-6}$, $2.5\times 10^{-7}$, and $2.5\times 10^{-8}~M_\odot$, the timescales that three systems evolve to the current orbital period of WX Cen are $11.4971$, $11.4979$, and $11.5102~\rm Gyr$, respectively. At $P = 0.417~\rm days$, three systems are experiencing a rapid orbital decay. A heavier CB disk can produce a steeper slope at the current orbital period of WX Cen, which corresponds to a larger $\dot{P}$. Compared with Figure 2, our detailed stellar evolution model requires a relatively heavy CB disk. This is because the mass transfer produces an orbital expansion effect when the mass is transferred from the less massive donor star to the more massive WD. To compensate for the influence of the orbital expansion effect, it requires a heavier CB disk to produce the observed $\dot{P}$ of WX Cen.

\begin{figure*}
\centering
\includegraphics[width=1.5\linewidth,trim={0 0 0 0},clip]{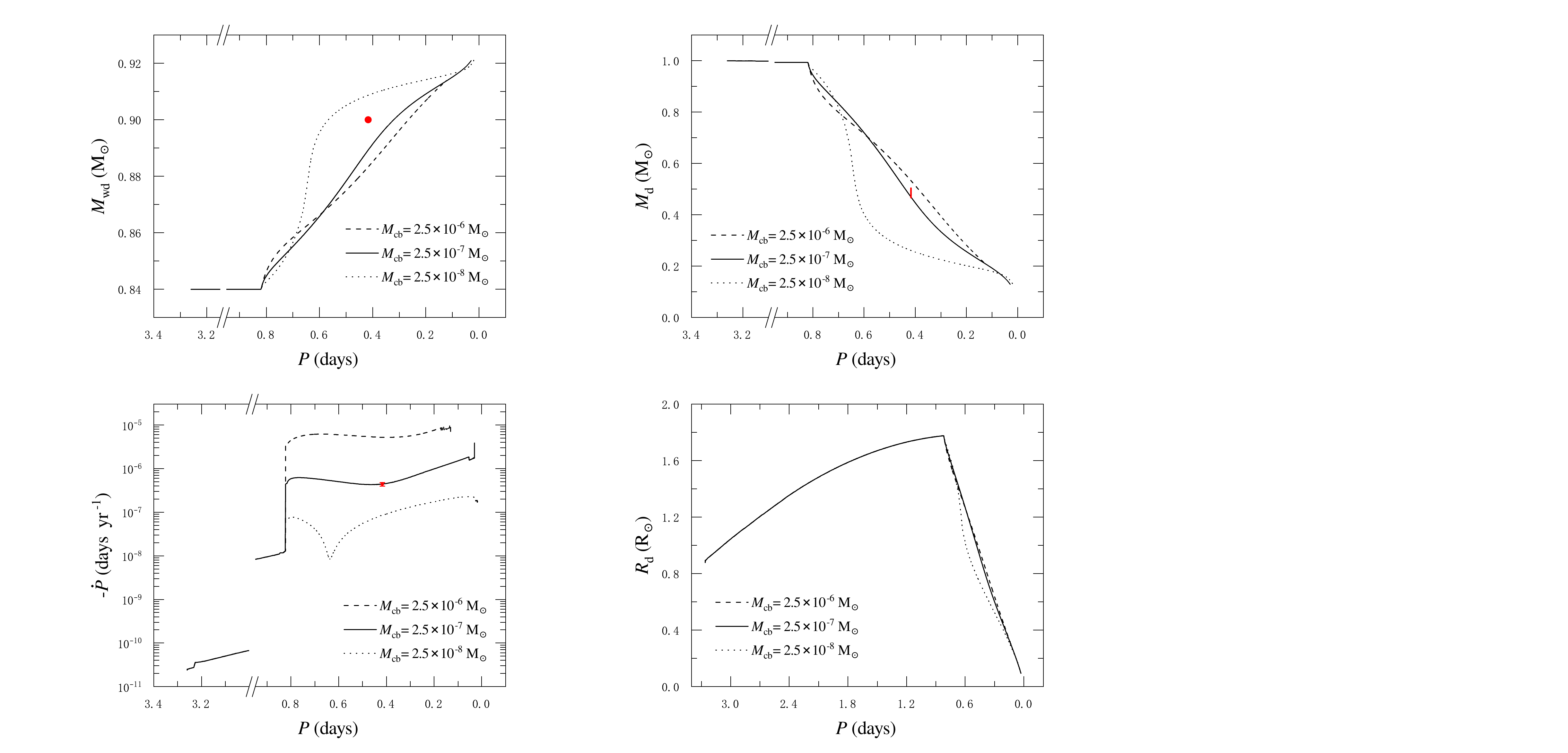}
\caption{Same as in Figure \ref{fig:3}, but for WD mass vs. orbital period (upper left panel),  donor-star mass vs. orbital period (upper right panel), orbital-period derivative vs. orbital period (bottom left panel), and donor-star radius vs. orbital period (bottom right panel) diagrams. The red solid circles and vertical line represent the observed and constrained parameters of WX Cen.} 
\label{fig:4}
\end{figure*}

The same as Figure \ref{fig:3}, Figure \ref{fig:4} compares our best model with the observed and our constrained parameters of WX Cen. In the early stage, donor stars continuously expand due to long-term nuclear evolution. At $P=0.82~\rm days$, donor stars fill their Roche lobes. Subsequently, donor stars begin to contract with the rapid orbital decay in semi-detached systems. Our simulated results are approximately consistent with the observed and constrained parameters including the orbital period derivative, WD mass, and donor-star mass of WX Cen in the current orbital period. At $P=0.417~\rm days$, the WD binary with $M_{\rm cb}=2.5 \times 10^{-7}~M_\odot$ has an orbital period derivative of $\dot{P}\approx-4.0\times10^{-7}~\rm days\,yr^{-1}$. The CB-disk mass plays a decisive role in influencing the orbital period derivative. A heavier CB disk can induce a higher orbital period derivative in the current state of WX Cen. 
 
\begin{figure}
\centering
\includegraphics[width=1.15\linewidth,trim={0 0 0 0},clip]{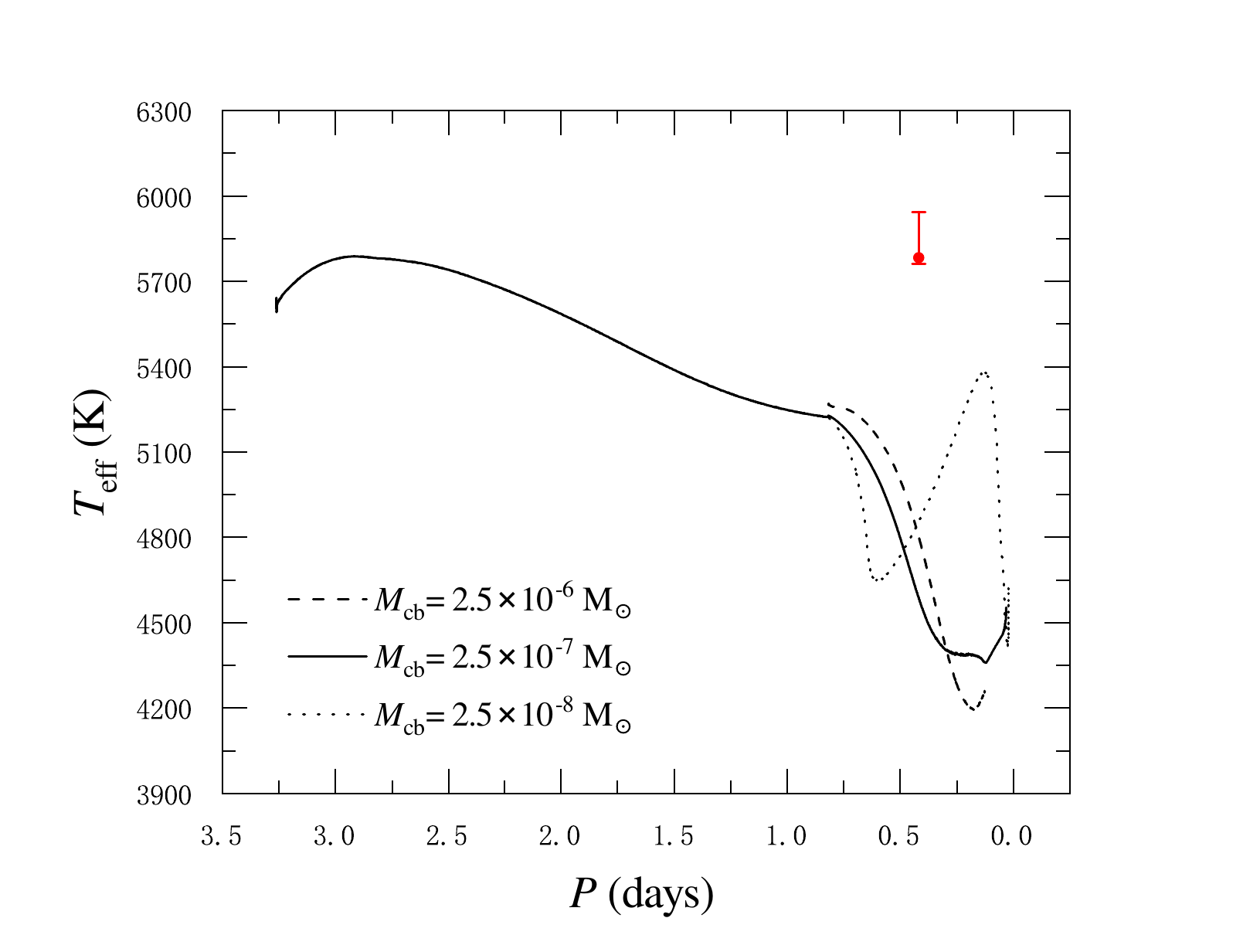}
\caption{Same as in Figure \ref{fig:3}, but for the donor star's effective temperature vs. orbital period diagram. The red solid circle with error bars represents the observed parameters of WX Cen.} 
\label{fig:5}
\end{figure}

Figure \ref{fig:5} depicts the evolution of donor stars in the effective temperature versus orbital period. At $P = 0.417~\rm days$, our simulated donor stars' effective temperatures are $4600-4900~\rm K$, which are much lower than the observed value ($5782.5^{+161.9}_{-21.5}~\rm K$) of WX Cen \citep{GaiaDR3-2022}. For a donor star with a mass of $1.0~M_\odot$, it has an effective temperature of $5600-5800~\rm K$ in its MS stage, which is similar to the Sun. However, it is unusual for a low-mass ($\sim 0.5~M_\odot$) donor star to have such a high effective temperature, which could come from the emission of a surrounding CB disk (see also section 6.1).

\begin{figure}
\centering
\includegraphics[width=1.0\linewidth,trim={0 0 0 0},clip]{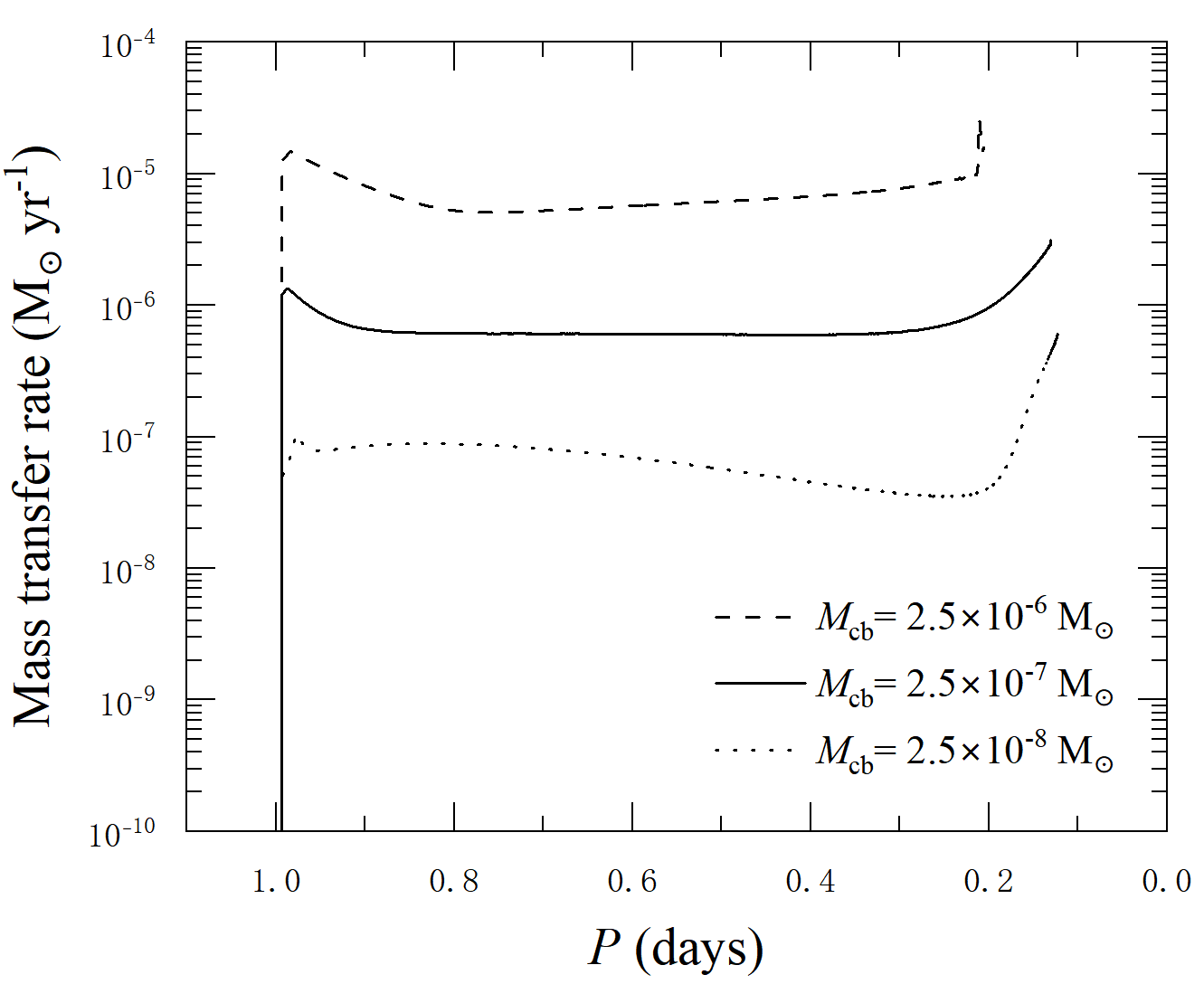}
\caption{Same as in Figure \ref{fig:3}, but for the mass transfer rate vs. donor-star mass diagram.}
\label{fig:6}
\end{figure}

\begin{figure}
\centering
\includegraphics[width=1.15\linewidth,trim={0 0 0 0},clip]{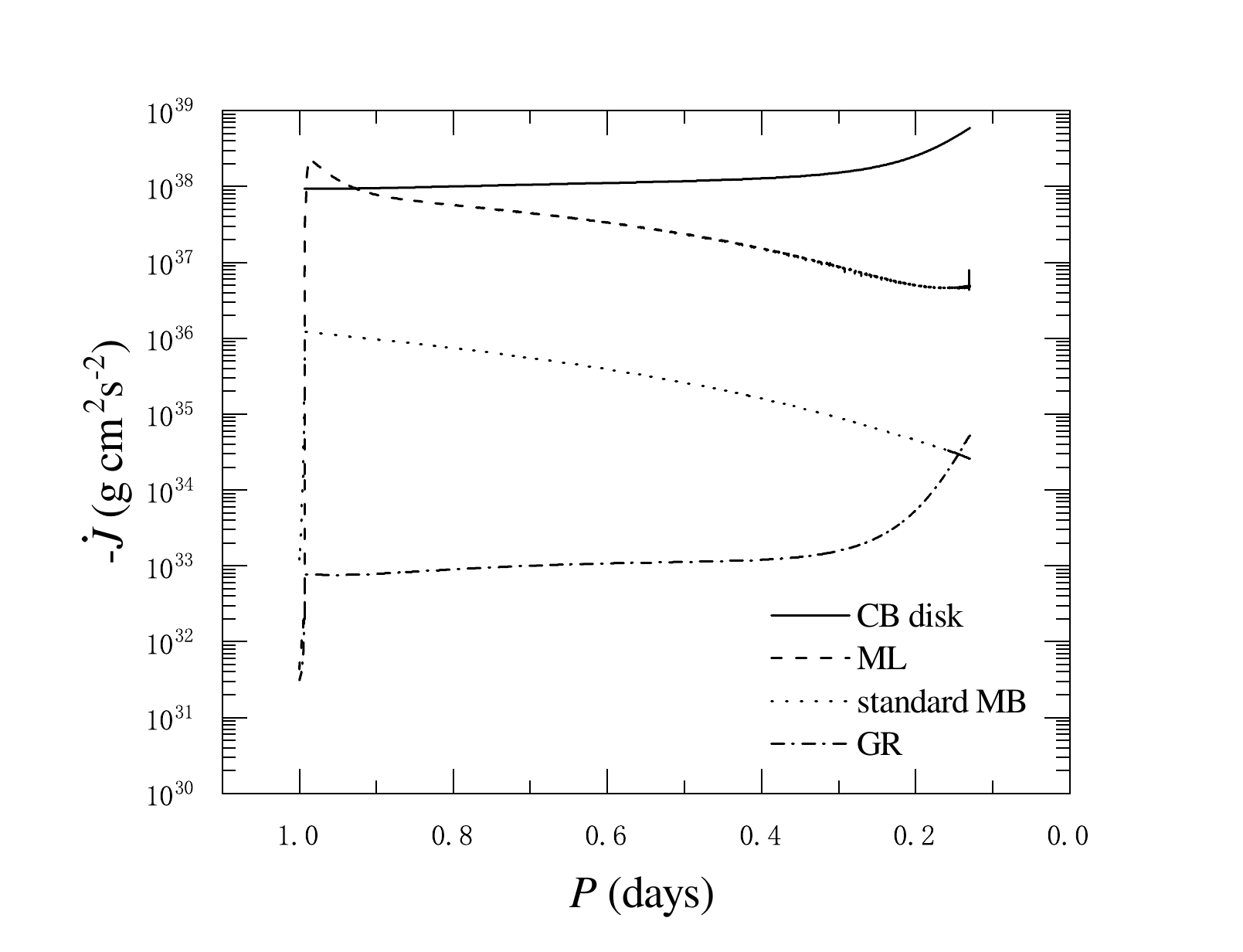}
\caption{Evolution of a WD binary with $M_{\rm wd,i}=0.84~M_\odot$, $M_{\rm d,i}=1.0~M_\odot$, and $P_{\rm i}=3.26~\rm days$ when $M_{\rm cb}=2.5\times10^{-7}~M_\odot$ in AML rates vs. donor-star masses diagram. The solid, dashed, dotted, and dashed-dotted curves correspond to the rates of AML due to a surrounding CB disk, mass loss, standard MB prescription, and gravitational wave radiation, respectively.} 
\label{fig:7}
\end{figure}

Figure \ref{fig:6} illustrates the evolution of the WD binary in the mass-transfer rate versus donor-star mass diagram. One can see that a heavier CB disk results in a more efficient AML and a higher mass-transfer rate. When $M_{\rm cb}=2.5\times10^{-7}~M_\odot$, the mass-transfer rate is $5.3\times10^{-7}~M_\odot\rm yr^{-1}$ at the current donor-star mass of WX Cen. This mass-transfer rate is much higher than the one ($\sim10^{-7}~M_\odot\rm yr^{-1}$) which can trigger a stable hydrogen burning process on the surface of the WD in WX Cen \citep{kaha1997}. If the accreted material can effectively accumulate on the surface of the WD, the accreting WD could reach the Chandrasekhar limit and trigger a Type Ia supernova in a timescale of $5~\rm Myr$ \citep{oliv04,Qian2013}. This is consistent with the prediction that SSSs open a \sout{new} channel to Type Ia supernovae \citep{li97,Hachisu1999}. The CB-disk mass plays a vital role in influencing the mass-transfer rate. A light CB disk ($M_{\rm cb}=2.5\times10^{-8}~M_\odot$) can only lead to a relatively low mass-transfer rate smaller than $10^{-7}~M_\odot\rm yr^{-1}$.

In Figure \ref{fig:7}, we summarize the rate of AML by different mechanisms in the AML rates versus donor-star masses diagram. It is negligible for the gravitational radiation, which has an AML rate that is $4-5$ orders of magnitude smaller than that by mass transfer and the surrounding CB disk. The rate of AML by the standard MB mechanism is two orders of magnitude smaller than that given by the mass loss. In the current orbital period ($0.417~\rm days$), the AML rate due to mass loss is approximately an order of magnitude smaller than that of the surrounding CB disk. As a consequence, the observed $\dot{P}$ is mainly contributed by the AML by the CB disk. Considering the weak dependence of $\dot{P}$ on mass transfer, our simulated $\dot{P}$ remains a certain degree of reliability even if we take a constant accretion efficiency in the numerical calculations.

\section{Discussion}
\label{sec:discussion}
\subsection{Detectability of the CB disk}
A direct detection of the CB disk surrounding WX Cen can confirm our proposed model. According to the torque ($T_{\rm cb}$) of the CB disk exerting on the binary, we can estimate its luminosity as $L_{\rm cb} = \Omega T_{\rm cb}=-2\pi \dot{J}_{\rm cb}/P$ \citep{Spruit2001}. In Case B (taking $M_{\rm d}=0.5~M_\odot$), the current rate of AML by the CB disk with $M_{\rm cb}=2.5\times10^{-7}~M_\odot$ is $\dot{J}_{\rm cb}=-1.2\times10^{38}~\rm g\,cm^2s^{-2}$, hence $L_{\rm cb}=2.1\times10^{34}~\rm erg\,s^{-1}$. Because the surface density and shear rate decrease with radius, the luminosity of the CB disk is contributed by a zone near its inner edge \citep{Spruit2001}. For WX Cen, the effective emitting area can be estimated to be $A\sim\pi r_{\rm in}^2=3.0\times10^{23}~\rm cm^2$. According to the Stefan-Boltzmann law, the effective temperature at the inner edge of the CB disk can be derived to be $T_{\rm eff}=[L_{\rm cb}/(\sigma A)]^{1/4}\approx5900~\rm K$, which is close to the effective temperature ($5782.5^{+161.9}_{-21.5}~\rm K$) detected in WX Cen \citep{GaiaDR3-2022}. 

Since the effective temperature of the CB disk decreases with increasing radius, the outer disk should produce mid-infrared emission. \citet{muno06} found that the excess mid-infrared emission originates from regions 2 times larger than the orbital separations of black hole LMXBs A0620-00 and XTE J1118+480, and about 4 times larger than their accretion disks. As a consequence, they proposed that the mid-infrared emission arises from surrounding CB disks with a mass of $\sim10^{-9}~M_\odot$. Subsequently, \cite{wang14} identified that A0620-00 and XTE J1118+480 are candidate systems with CB disks employing the Wide-field Infrared Survey Explorer observation. Coincidentally, \citet{chen2019} confirmed that the surrounding CB disks with $M_{\rm cb}\sim10^{-9}~M_\odot$ can be responsible for the rapid orbital decay ($\dot{P}\sim10^{-8}~\rm days~yr^{-1}$) detected in A0620-00 and XTE J1118+480. Therefore, the detection of the excess mid-infrared emission from WX Cen can help us to confirm or rule out the proposed CB disk scenario.

\subsection{Other possible mechanisms}
\subsubsection{Magnetic activity cycles}
If the rapid orbital decay observed in WX Cen is a short-term phenomenon, it may originate from orbital period modulations induced by the gravitational coupling of the binary orbit to the deformation of the donor star with magnetic active cycles \citep{appl92}. Magnetic activity can transfer angular momentum to different convective zones of the star to cause a deformation. The change in the moment of inertia induces a variation in the spin period of the star, naturally resulting in an orbital period change due to the tidal locking \citep{appl92}. \citet{yang14} proposed that magnetic activity cycles of the secondary star can account for the orbital decay observed in the Algol system AF Gem. If the magnetic activity cycles can work for WX Cen, it implies that its donor star has a strong magnetic field of $\sim1000~\rm G$. A long-term detection of WX Cen could confirm or rule out this mechanism.
\subsubsection{Dynamical friction}
Dynamical friction within an expanding nova shell could also drive a rapid orbital decay in WD binaries \citep{shen15}. When the accretion rate is less than $10^{-6}~ M_\odot \rm yr^{-1}$ in a WD binary, the unstable burning of the accreted helium triggers a helium flash, eventually resulting in a nova shell surrounding the binary \citep{shen09}. Similar to the ejection process of a common envelope \citep{ivan13,post14}, the ejection of the WD's expanding nova shell causes the binary separation to decay. Assuming that the energy to eject the shell comes from the binding energy of the orbit, one can obtain a change of orbital separation \citep[see also equation 2 of][]{shen15}.  Differentiating Kepler’s third law ($G(M_{\rm wd}+M_{\rm d})/a^{3}=4\pi^{2}/P^{2}$) and ignoring the orbital decay due to the mass loss (mass loss cannot produce the orbital decay observed in WX Cen, see also Section 2), it yields $\dot{P}/P=3\dot{a}/(2a)$. Therefore, the dynamical friction can produce an orbital period derivative as
\begin{equation}
\dot{P}_{\rm df} = -\frac{\dot{M}_{\rm ej}}{M_{\rm wd}}\left[\frac{3(1+q)f(q)}{q}+\frac{3}{2}\right]P,    
\end{equation}
where $\dot{M}_{\rm ej}$ is the ejecta-loss rate and $f(q)$ is the function of the mass ratio \citep[see also equation 1 of][]{shen15}. 

Taking into account Case B (taking $M_{\rm d}=0.5~M_\odot$) in our constraint for WX Cen, $q=0.56$, we can derive $f(q)=1.94$. If $\dot{M}_{\rm ej}=5\times10^{-8}~M_\odot\,\rm yr^{-1}$, the dynamical friction can produce an orbital period derivative of $\dot{P}_{\rm df}\approx4.1\times 10^{-7}~\rm days\,yr^{-1}$. Similar to a surrounding CB disk, it is possible to produce a high mass-transfer rate of $\sim10^{-7}~M_\odot\,\rm yr^{-1}$ because the dynamical friction provides an efficient AML mechanism. As a result, a mass loss is expected from the surface of the accreting WD at a rate $\sim9\times10^{-8}~M_\odot\,\rm yr^{-1}$ when the accretion efficiency is 0.1, and an ejecta-loss rate of $\sim5\times10^{-8}~M_\odot\,\rm yr^{-1}$ is highly feasible. Therefore, the dynamical friction could also be a potential mechanism that resulted in the rapid orbital decay of WX Cen.

\section{Summary}
\label{sec:summary}
Recently, WX Cen, a candidate of compact binary SSSs,  was detected to be experiencing a rapid orbital decay at a rate of $\dot{P} = -(4.4\pm0.4) \times 10^{-7}~ \rm days\ \rm{yr}^{-1}$ \citep{zang23}. According to the mass function and optical eclipses, the donor-star mass can be constrained to be $0.41-0.44$,  $0.47-0.50$, and $0.55-0.59~M_\odot$ when the WD mass is $0.7$, $0.9$, and $1.2~M_\odot$, respectively. Our calculations find that this orbital period derivative cannot be produced by the AML due to mass loss during the accretion of the WD, MB mechanisms including the standard MB, the CARB MB, and the anomalous MB prescriptions, implying that an efficient AML mechanism exists during the evolution of WX Cen.  

In this work, we propose that a surrounding CB disk could be responsible for the rapid orbital decay of WX Cen. When $\alpha = 0.1$, $H/R = 0.1$, and $R/a = 4.12$, the predicted $\dot{P}$ can reach $\dot{P}\sim-4.0\times10^{-7}~ \rm days~\rm{yr}^{-1}$ for the tidal torque originating from the resonant interaction between the binary and a CB disk with a mass of $10^{-7}~M_\odot$. We also perform a detailed binary evolution model for the formation of WX Cen. When $M_{\rm cb}=2.5\times 10^{-7}~M_\odot$, a WD binary with $M_{\rm wd,i} = 0.84 ~M_\odot$, $M_{\rm d,i} = 1.0~M_\odot$, and $P_{\rm i} = 3.26~\rm days$ can evolve into a WX Cen-like system, which has an orbital period and orbital period derivative same to WX Cen, and donor-star and WD masses similar to our constrained Case B. In the current state of WX Cen, our simulated mass-transfer rate is $5.3\times10^{-7}~M_\odot\rm yr^{-1}$, which is much higher than the critical value ($\sim10^{-7}~M_\odot\rm yr^{-1}$) that can trigger a stable hydrogen burning process on the surface of the WD.

%\begin{acknowledgments}
%We thank the referee for a very careful reading and constructive comments that have led to the improvement of the manuscript. This work was partially funded by the National Natural Science Foundation of China (Grant No. 12273014) and the Natural Science Foundation of Shandong Province (Grant No. ZR2025MS43).
%\end{acknowledgments}

\bibliography{ref}

@ARTICLE{zang23,
       author = {{Zang}, Lei and {Qian}, Shengbang and {Fern{\'a}ndez-Laj{\'u}s}, Eduardo},
        title = "{The Evolution of the Supersoft X-Ray Source WX Cen Dominated by Magnetic Wind}",
      journal = {\apj},
         year = 2023,
        month = feb,
       volume = {944},
       number = {1},
          eid = {97},
        pages = {97},
          doi = {10.3847/1538-4357/acb52f},
       adsurl = {https://ui.adsabs.harvard.edu/abs/2023ApJ...944...97Z}
}

@ARTICLE{oome20,
       author = {{Oomen}, Glenn-Michael and {Pols}, Onno and {Van Winckel}, Hans and {Nelemans}, Gijs},
        title = "{Disc-binary interactions in depleted post-AGB binaries}",
      journal = {\aap},
         year = 2020,
        month = oct,
       volume = {642},
          eid = {A234},
        pages = {A234},
          doi = {10.1051/0004-6361/202038341},
archivePrefix = {arXiv},
       eprint = {2008.08097},
 primaryClass = {astro-ph.SR},
       adsurl = {https://ui.adsabs.harvard.edu/abs/2020A&A...642A.234O}
}

@ARTICLE{reim75,
       author = {{Reimers}, D.},
        title = "{Circumstellar absorption lines and mass loss from red giants.}",
      journal = {Memoires of the Societe Royale des Sciences de Liege},
         year = 1975,
        month = jan,
       volume = {8},
        pages = {369-382},
       adsurl = {https://ui.adsabs.harvard.edu/abs/1975MSRSL...8..369R}
}

@BOOK{grei96,
       author = {{Greiner}, Jochen},
        title = "{Supersoft X-Ray Sources}",
         year = 1996,
       volume = {472},
          doi = {10.1007/BFb0102238},
       adsurl = {https://ui.adsabs.harvard.edu/abs/1996LNP...472.....G}
}

@ARTICLE{egge68,
       author = {{Eggen}, O.~J. and {Freeman}, Kenneth C. and {Sandage}, Allan},
        title = "{On the Optical Identification of the X-Ray Source CEN XR-2 as WX CEN}",
      journal = {\apjl},
         year = 1968,
        month = oct,
       volume = {154},
        pages = {L27},
          doi = {10.1086/180261},
       adsurl = {https://ui.adsabs.harvard.edu/abs/1968ApJ...154L..27E}
}

@ARTICLE{long81,
       author = {{Long}, K.~S. and {Helfand}, D.~J. and {Grabelsky}, D.~A.},
        title = "{A soft X-ray study of the Large Magellanic Cloud.}",
      journal = {\apj},
         year = 1981,
        month = sep,
       volume = {248},
        pages = {925-944},
          doi = {10.1086/159222},
       adsurl = {https://ui.adsabs.harvard.edu/abs/1981ApJ...248..925L}
}

@ARTICLE{li97,
       author = {{Li}, X.-D. and {van den Heuvel}, E.~P.~J.},
        title = "{Evolution of white dwarf binaries: supersoft X-ray sources and progenitors of type IA supernovae.}",
      journal = {\aap},
         year = 1997,
        month = jun,
       volume = {322},
        pages = {L9-L12},
       adsurl = {https://ui.adsabs.harvard.edu/abs/1997A&A...322L...9L}
}

@ARTICLE{huch81,
       author = {{van der Hucht}, K.~A. and {Conti}, P.~S. and {Lundstrom}, I. and {Stenholm}, B.},
        title = "{The Sixth Catalogue of Galactic Wolf-Rayet Stars - Their Past and Present}",
      journal = {\ssr},
         year = 1981,
        month = sep,
       volume = {28},
       number = {3},
        pages = {227-306},
          doi = {10.1007/BF00173260},
       adsurl = {https://ui.adsabs.harvard.edu/abs/1981SSRv...28..227V}
}

@BOOK{taur23,
       author = {{Tauris}, Thomas M. and {van den Heuvel}, Edward P.~J.},
        title = "{Physics of Binary Star Evolution. From Stars to X-ray Binaries and Gravitational Wave Sources}",
         year = 2023,
          doi = {10.48550/arXiv.2305.09388},
       adsurl = {https://ui.adsabs.harvard.edu/abs/2023pbse.book.....T}
}

@ARTICLE{muno06,
       author = {{Muno}, Michael P. and {Mauerhan}, Jon},
        title = "{Mid-Infrared Emission from Dust around Quiescent Low-Mass X-Ray Binaries}",
      journal = {\apjl},
         year = 2006,
        month = sep,
       volume = {648},
       number = {2},
        pages = {L135-L138},
          doi = {10.1086/507990},
archivePrefix = {arXiv},
       eprint = {astro-ph/0607083},
 primaryClass = {astro-ph},
       adsurl = {https://ui.adsabs.harvard.edu/abs/2006ApJ...648L.135M}
}

@ARTICLE{yang14,
       author = {{Yang}, Yuan-Gui and {Yang}, Ying and {Li}, Shu-Zheng},
        title = "{Photometric Properties for Selected Algol-type Binaries. VIII. The Triple Systems DI Peg and AF Gem Revisited}",
      journal = {\aj},
         year = 2014,
        month = jun,
       volume = {147},
       number = {6},
          eid = {145},
        pages = {145},
          doi = {10.1088/0004-6256/147/6/145},
       adsurl = {https://ui.adsabs.harvard.edu/abs/2014AJ....147..145Y}
}

@ARTICLE{shen15,
       author = {{Shen}, Ken J.},
        title = "{Every Interacting Double White Dwarf Binary May Merge}",
      journal = {\apjl},
         year = 2015,
        month = may,
       volume = {805},
       number = {1},
          eid = {L6},
        pages = {L6},
          doi = {10.1088/2041-8205/805/1/L6},
archivePrefix = {arXiv},
       eprint = {1502.05052},
 primaryClass = {astro-ph.SR},
       adsurl = {https://ui.adsabs.harvard.edu/abs/2015ApJ...805L...6S}
}

@ARTICLE{post14,
       author = {{Postnov}, Konstantin A. and {Yungelson}, Lev R.},
        title = "{The Evolution of Compact Binary Star Systems}",
      journal = {Living Reviews in Relativity},
         year = 2014,
        month = dec,
       volume = {17},
       number = {1},
          eid = {3},
        pages = {3},
          doi = {10.12942/lrr-2014-3},
archivePrefix = {arXiv},
       eprint = {1403.4754},
 primaryClass = {astro-ph.HE},
       adsurl = {https://ui.adsabs.harvard.edu/abs/2014LRR....17....3P}
}

@ARTICLE{ivan13,
       author = {{Ivanova}, N. and {Justham}, S. and {Chen}, X. and {De Marco}, O. and {Fryer}, C.~L. and {Gaburov}, E. and {Ge}, H. and {Glebbeek}, E. and {Han}, Z. and {Li}, X.-D. and {Lu}, G. and {Marsh}, T. and {Podsiadlowski}, P. and {Potter}, A. and {Soker}, N. and {Taam}, R. and {Tauris}, T.~M. and {van den Heuvel}, E.~P.~J. and {Webbink}, R.~F.},
        title = "{Common envelope evolution: where we stand and how we can move forward}",
      journal = {\aapr},
         year = 2013,
        month = feb,
       volume = {21},
          eid = {59},
        pages = {59},
          doi = {10.1007/s00159-013-0059-2},
archivePrefix = {arXiv},
       eprint = {1209.4302},
 primaryClass = {astro-ph.HE},
       adsurl = {https://ui.adsabs.harvard.edu/abs/2013A&ARv..21...59I}
}

@ARTICLE{shen09,
       author = {{Shen}, Ken J. and {Bildsten}, Lars},
        title = "{Unstable Helium Shell Burning on Accreting White Dwarfs}",
      journal = {\apj},
         year = 2009,
        month = jul,
       volume = {699},
       number = {2},
        pages = {1365-1373},
          doi = {10.1088/0004-637X/699/2/1365},
archivePrefix = {arXiv},
       eprint = {0903.0654},
 primaryClass = {astro-ph.HE},
       adsurl = {https://ui.adsabs.harvard.edu/abs/2009ApJ...699.1365S}
}

@ARTICLE{liu19,
       author = {{Liu}, Wei-Min and {Li}, Xiang-Dong},
        title = "{Can the Friction of the Nova Envelope Account for the Extra Angular Momentum Loss in Cataclysmic Variables?}",
      journal = {\apj},
         year = 2019,
        month = jan,
       volume = {870},
       number = {1},
          eid = {22},
        pages = {22},
          doi = {10.3847/1538-4357/aaf19f},
archivePrefix = {arXiv},
       eprint = {1811.08648},
 primaryClass = {astro-ph.SR},
       adsurl = {https://ui.adsabs.harvard.edu/abs/2019ApJ...870...22L}
}

@ARTICLE{appl92,
       author = {{Applegate}, James H.},
        title = "{A Mechanism for Orbital Period Modulation in Close Binaries}",
      journal = {\apj},
         year = 1992,
        month = feb,
       volume = {385},
        pages = {621},
          doi = {10.1086/170967},
       adsurl = {https://ui.adsabs.harvard.edu/abs/1992ApJ...385..621A}
}

@ARTICLE{daiz1995,
       author = {{Diaz}, M.~P. and {Steiner}, J.~E.},
        title = "{The Nova-like Variable WX Centauri and the V Sagittae Phenomenon}",
      journal = {\aj},
         year = 1995,
        month = oct,
       volume = {110},
        pages = {1816},
          doi = {10.1086/117653},
       adsurl = {https://ui.adsabs.harvard.edu/abs/1995AJ....110.1816D}
}

@ARTICLE{oliv04,
       author = {{Oliveira}, A.~S. and {Steiner}, J.~E.},
        title = "{WX Cen ({\ensuremath{\equiv}} WR 48c) - a possible Type Ia supernova progenitor}",
      journal = {\mnras},
         year = 2004,
        month = jun,
       volume = {351},
       number = {2},
        pages = {685-693},
          doi = {10.1111/j.1365-2966.2004.07817.x},
archivePrefix = {arXiv},
       eprint = {astro-ph/0403281},
 primaryClass = {astro-ph},
       adsurl = {https://ui.adsabs.harvard.edu/abs/2004MNRAS.351..685O}
}

@ARTICLE{patt98,
       author = {{Patterson}, Joseph and {Kemp}, Jonathan and {Shambrook}, Anouk and {Thorstensen}, John R. and {Skillman}, David R. and {Gunn}, Jerry and {Jensen}, Lasse and {Vanmunster}, Tonny and {Shugarov}, Sergei and {Mattei}, Janet A. and {Shahbaz}, Tariq and {Novak}, Rudolf},
        title = "{Two Galactic Supersoft X-Ray Binaries: V Sagittae and T Pyxidis}",
      journal = {\pasp},
         year = 1998,
        month = apr,
       volume = {110},
       number = {746},
        pages = {380-395},
          doi = {10.1086/316147},
       adsurl = {https://ui.adsabs.harvard.edu/abs/1998PASP..110..380P}
}

@ARTICLE{xu18,
       author = {{Xu}, Xiao-Tian and {Li}, Xiang-Dong},
        title = "{A Circumbinary Disk Model for the Rapid Orbital Shrinkage in Black Hole Low-mass X-Ray Binaries}",
      journal = {\apj},
         year = 2018,
        month = may,
       volume = {859},
       number = {1},
          eid = {46},
        pages = {46},
          doi = {10.3847/1538-4357/aabe91},
archivePrefix = {arXiv},
       eprint = {1804.07914},
 primaryClass = {astro-ph.HE},
       adsurl = {https://ui.adsabs.harvard.edu/abs/2018ApJ...859...46X}
}

@ARTICLE{heuv1992,
       author = {{van den Heuvel}, E.~P.~J. and {Bhattacharya}, D. and {Nomoto}, K. and {Rappaport}, S.~A.},
        title = "{Accreting white dwarf models for CAL 83, CAL 87 and other ultrasoft X-ray sources in the LMC.}",
      journal = {\aap},
         year = 1992,
        month = aug,
       volume = {262},
        pages = {97-105},
       adsurl = {https://ui.adsabs.harvard.edu/abs/1992A&A...262...97V}
}

@ARTICLE{kaha1997,
       author = {{Kahabka}, P. and {van den Heuvel}, E.~P.~J.},
        title = "{Luminous Supersoft X-Ray Sources}",
      journal = {\araa},
         year = 1997,
        month = jan,
       volume = {35},
        pages = {69-100},
          doi = {10.1146/annurev.astro.35.1.69},
       adsurl = {https://ui.adsabs.harvard.edu/abs/1997ARA&A..35...69K}
}

@ARTICLE{Qian2013,
       author = {{Qian}, S.-B. and {Shi}, G. and {Fern{\'a}ndez Laj{\'u}s}, E. and {Di Sisto}, R.~P. and {Zhu}, L.-Y. and {Liu}, L. and {Zhao}, E.-G. and {Li}, L.-J.},
        title = "{Is WX Cen a Possible Type Ia Supernova Progenitor with Wind-driven Mass Transfer?}",
      journal = {\apjl},
         year = 2013,
        month = aug,
       volume = {772},
       number = {2},
          eid = {L18},
        pages = {L18},
          doi = {10.1088/2041-8205/772/2/L18},
       adsurl = {https://ui.adsabs.harvard.edu/abs/2013ApJ...772L..18Q}
}

@ARTICLE{Rappaport1983,
       author = {{Rappaport}, S. and {Verbunt}, F. and {Joss}, P.~C.},
        title = "{A new technique for calculations of binary stellar evolution application to magnetic braking.}",
      journal = {\apj},
         year = 1983,
        month = dec,
       volume = {275},
        pages = {713-731},
          doi = {10.1086/161569},
       adsurl = {https://ui.adsabs.harvard.edu/abs/1983ApJ...275..713R}
}

@ARTICLE{Verbunt1981,
       author = {{Verbunt}, F. and {Zwaan}, C.},
        title = "{Magnetic braking in low-mass X-ray binaries.}",
      journal = {\aap},
         year = 1981,
        month = jul,
       volume = {100},
        pages = {L7-L9},
       adsurl = {https://ui.adsabs.harvard.edu/abs/1981A&A...100L...7V}
}

@ARTICLE{Van2019,
       author = {{Van}, Kenny X. and {Ivanova}, Natalia},
        title = "{Evolving LMXBs: CARB Magnetic Braking}",
      journal = {\apjl},
         year = 2019,
        month = dec,
       volume = {886},
       number = {2},
          eid = {L31},
        pages = {L31},
          doi = {10.3847/2041-8213/ab571c},
archivePrefix = {arXiv},
       eprint = {1911.05790},
 primaryClass = {astro-ph.SR},
       adsurl = {https://ui.adsabs.harvard.edu/abs/2019ApJ...886L..31V}
}

@ARTICLE{tout88,
       author = {{Tout}, Christopher A. and {Eggleton}, Peter P.},
        title = "{Tidal enhancement by a binary companion of stellar winds from cool giants.}",
      journal = {\mnras},
         year = 1988,
        month = apr,
       volume = {231},
        pages = {823-831},
          doi = {10.1093/mnras/231.4.823},
       adsurl = {https://ui.adsabs.harvard.edu/abs/1988MNRAS.231..823T}
}

@ARTICLE{Victor2015,
       author = {{R{\'e}ville}, Victor and {Brun}, Allan Sacha and {Matt}, Sean P. and {Strugarek}, Antoine and {Pinto}, Rui F.},
        title = "{The Effect of Magnetic Topology on Thermally Driven Wind: Toward a General Formulation of the Braking Law}",
      journal = {\apj},
         year = 2015,
        month = jan,
       volume = {798},
       number = {2},
          eid = {116},
        pages = {116},
          doi = {10.1088/0004-637X/798/2/116},
archivePrefix = {arXiv},
       eprint = {1410.8746},
 primaryClass = {astro-ph.SR},
       adsurl = {https://ui.adsabs.harvard.edu/abs/2015ApJ...798..116R}
}

@ARTICLE{Justham2006,
       author = {{Justham}, Stephen and {Rappaport}, Saul and {Podsiadlowski}, Philipp},
        title = "{Magnetic braking of Ap/Bp stars: application to compact black-hole X-ray binaries}",
      journal = {\mnras},
         year = 2006,
        month = mar,
       volume = {366},
       number = {4},
        pages = {1415-1423},
          doi = {10.1111/j.1365-2966.2005.09907.x},
archivePrefix = {arXiv},
       eprint = {astro-ph/0511760},
 primaryClass = {astro-ph},
       adsurl = {https://ui.adsabs.harvard.edu/abs/2006MNRAS.366.1415J}
}

@ARTICLE{den1973,
       author = {{van den Heuvel}, E.~P.~J. and {De Loore}, C.},
        title = "{The nature of X-ray binaries III. Evolution of massive close binaries with one collapsed component - with a possible application to Cygnus X-3.}",
      journal = {\aap},
         year = 1973,
        month = jun,
       volume = {25},
        pages = {387},
       adsurl = {https://ui.adsabs.harvard.edu/abs/1973A&A....25..387V}
}

@INPROCEEDINGS{den1994,
       author = {{van den Heuvel}, E.~P.~J.},
        title = "{Interacting binaries: topics in close binary evolution.}",
    booktitle = {Saas-Fee Advanced Course 22: Interacting Binaries},
         year = 1994,
       editor = {{Shore}, S.~N. and {Livio}, M. and {van den Heuvel}, Edward P.~J. and {Nussbaumer}, H. and {Orr}, Astrid},
        month = jan,
        pages = {263-474},
       adsurl = {https://ui.adsabs.harvard.edu/abs/1994inbi.conf..263V}
}

@ARTICLE{Chen2006,
       author = {{Chen}, Wen-Cong and {Li}, Xiang-Dong},
        title = "{Evolution of black hole intermediate-mass X-ray binaries: the influence of a circumbinary disc}",
      journal = {\mnras},
         year = 2006,
        month = nov,
       volume = {373},
       number = {1},
        pages = {305-310},
          doi = {10.1111/j.1365-2966.2006.11032.x},
archivePrefix = {arXiv},
       eprint = {astro-ph/0609093},
 primaryClass = {astro-ph},
       adsurl = {https://ui.adsabs.harvard.edu/abs/2006MNRAS.373..305C}
}

@ARTICLE{wang14,
       author = {{Wang}, Xuebing and {Wang}, Zhongxiang},
        title = "{WISE Detection of the Galactic Low-mass X-Ray Binaries}",
      journal = {\apj},
         year = 2014,
        month = jun,
       volume = {788},
       number = {2},
          eid = {184},
        pages = {184},
          doi = {10.1088/0004-637X/788/2/184},
archivePrefix = {arXiv},
       eprint = {1404.3472},
 primaryClass = {astro-ph.HE},
       adsurl = {https://ui.adsabs.harvard.edu/abs/2014ApJ...788..184W}
}

@ARTICLE{chen2017,
       author = {{Chen}, Wen-Cong and {Podsiadlowski}, Philipp},
        title = "{Rapid Orbital Decay in Detached Binaries: Evidence for Circumbinary Disks}",
      journal = {\apjl},
         year = 2017,
        month = mar,
       volume = {837},
       number = {2},
          eid = {L19},
        pages = {L19},
          doi = {10.3847/2041-8213/aa624a},
archivePrefix = {arXiv},
       eprint = {1702.06311},
 primaryClass = {astro-ph.SR},
       adsurl = {https://ui.adsabs.harvard.edu/abs/2017ApJ...837L..19C}
}

@ARTICLE{Spruit2001,
       author = {{Spruit}, H.~C. and {Taam}, Ronald E.},
        title = "{Circumbinary Disks and Cataclysmic Variable Evolution}",
      journal = {\apj},
         year = 2001,
        month = feb,
       volume = {548},
       number = {2},
        pages = {900-907},
          doi = {10.1086/319030},
archivePrefix = {arXiv},
       eprint = {astro-ph/0010194},
 primaryClass = {astro-ph},
       adsurl = {https://ui.adsabs.harvard.edu/abs/2001ApJ...548..900S}
}

@ARTICLE{Taam2001,
       author = {{Taam}, Ronald E. and {Spruit}, H.~C.},
        title = "{The Evolution of Cataclysmic Variable Binary Systems with Circumbinary Disks}",
      journal = {\apj},
         year = 2001,
        month = nov,
       volume = {561},
       number = {1},
        pages = {329-336},
          doi = {10.1086/322331},
       adsurl = {https://ui.adsabs.harvard.edu/abs/2001ApJ...561..329T}
}

@article{Han2018,
doi = {10.3847/1538-4357/aae64d},
url = {https://doi.org/10.3847/1538-4357/aae64d},
year = {2018},
month = {nov},
publisher = {The American Astronomical Society},
volume = {868},
number = {1},
pages = {53},
author = {Han, Z.-T. and Qian, S.-B. and Zhu, L.-Y. and Zhi, Q.-J. and Dong, A.-J. and Soonthornthum, B. and Poshyachinda, S. and Sarotsakulchai, T. and Fang, X.-H. and Wang, Q.-S. and Voloshina, Irina},
title = {DE CVn: An Eclipsing Post-common Envelope Binary with a Circumbinary Disk and a Giant Planet},
journal = {The Astrophysical Journal}

}

@ARTICLE{chen.el2006,
       author = {{Chen}, Wen-Cong and {Li}, Xiang-Dong and {Qian}, Sheng-Bang},
        title = "{Orbital Evolution of Algol Binaries with a Circumbinary Disk}",
      journal = {\apj},
         year = 2006,
        month = oct,
       volume = {649},
       number = {2},
        pages = {973-978},
          doi = {10.1086/506433},
archivePrefix = {arXiv},
       eprint = {astro-ph/0606081},
 primaryClass = {astro-ph},
       adsurl = {https://ui.adsabs.harvard.edu/abs/2006ApJ...649..973C}
}

@ARTICLE{chen2019,
       author = {{Chen}, Wen-Cong and {Podsiadlowski}, Philipp},
        title = "{Fast Orbital Shrinkage of Black Hole X-Ray Binaries Driven by Circumbinary Disks}",
      journal = {\apjl},
         year = 2019,
        month = may,
       volume = {876},
       number = {1},
          eid = {L11},
        pages = {L11},
          doi = {10.3847/2041-8213/ab1b44},
archivePrefix = {arXiv},
       eprint = {1904.09753},
 primaryClass = {astro-ph.HE},
       adsurl = {https://ui.adsabs.harvard.edu/abs/2019ApJ...876L..11C}
}

@ARTICLE{wei2023,
       author = {{Wei}, Na and {Jiang}, Long and {Chen}, Wen-Cong},
        title = "{Anomalous orbital expansion of the low-mass X-ray binary 2A 1822-371: the existence of a circumbinary disk?}",
      journal = {\aap},
         year = 2023,
        month = nov,
       volume = {679},
          eid = {A74},
        pages = {A74},
          doi = {10.1051/0004-6361/202346397},
archivePrefix = {arXiv},
       eprint = {2309.11529},
 primaryClass = {astro-ph.HE},
       adsurl = {https://ui.adsabs.harvard.edu/abs/2023A&A...679A..74W}
}

@article{Paxton_2011,
doi = {10.1088/0067-0049/192/1/3},
url = {https://doi.org/10.1088/0067-0049/192/1/3},
year = {2010},
month = {dec},
publisher = {The American Astronomical Society},
volume = {192},
number = {1},
pages = {3},
author = {Paxton, Bill and Bildsten, Lars and Dotter, Aaron and Herwig, Falk and Lesaffre, Pierre and Timmes, Frank},
title = {MODULES FOR EXPERIMENTS IN STELLAR ASTROPHYSICS (MESA)},
journal = {The Astrophysical Journal Supplement Series}
}

@article{Paxton_2013,
doi = {10.1088/0067-0049/208/1/4},
url = {https://doi.org/10.1088/0067-0049/208/1/4},
year = {2013},
month = {aug},
publisher = {The American Astronomical Society},
volume = {208},
number = {1},
pages = {4},
author = {Paxton, Bill and Cantiello, Matteo and Arras, Phil and Bildsten, Lars and Brown, Edward F. and Dotter, Aaron and Mankovich, Christopher and Montgomery, M. H. and Stello, Dennis and Timmes, F. X. and Townsend, Richard},
title = {MODULES FOR EXPERIMENTS IN STELLAR ASTROPHYSICS (MESA): PLANETS, OSCILLATIONS, ROTATION, AND MASSIVE STARS},
journal = {The Astrophysical Journal Supplement Series}
}

@article{Paxton_2015,
doi = {10.1088/0067-0049/220/1/15},
url = {https://doi.org/10.1088/0067-0049/220/1/15},
year = {2015},
month = {sep},
publisher = {The American Astronomical Society},
volume = {220},
number = {1},
pages = {15},
author = {Paxton, Bill and Marchant, Pablo and Schwab, Josiah and Bauer, Evan B. and Bildsten, Lars and Cantiello, Matteo and Dessart, Luc and Farmer, R. and Hu, H. and Langer, N. and Townsend, R. H. D. and Townsley, Dean M. and Timmes, F. X.},
title = {MODULES FOR EXPERIMENTS IN STELLAR ASTROPHYSICS (MESA): BINARIES, PULSATIONS, AND EXPLOSIONS},
journal = {The Astrophysical Journal Supplement Series}
}

@article{Paxton_2018,
doi = {10.3847/1538-4365/aaa5a8},
url = {https://doi.org/10.3847/1538-4365/aaa5a8},
year = {2018},
month = {feb},
publisher = {The American Astronomical Society},
volume = {234},
number = {2},
pages = {34},
author = {Paxton, Bill and Schwab, Josiah and Bauer, Evan B. and Bildsten, Lars and Blinnikov, Sergei and Duffell, Paul and Farmer, R. and Goldberg, Jared A. and Marchant, Pablo and Sorokina, Elena and Thoul, Anne and Townsend, Richard H. D. and Timmes, F. X.},
title = {Modules for Experiments in Stellar Astrophysics ({\mathtt{M}}{\mathtt{E}}{\mathtt{S}}{\mathtt{A}}): Convective Boundaries, Element Diffusion, and Massive Star Explosions},
journal = {The Astrophysical Journal Supplement Series}
}

@article{Paxton_2019,
doi = {10.3847/1538-4365/ab2241},
url = {https://doi.org/10.3847/1538-4365/ab2241},
year = {2019},
month = {jul},
publisher = {The American Astronomical Society},
volume = {243},
number = {1},
pages = {10},
author = {Paxton, Bill and Smolec, R. and Schwab, Josiah and Gautschy, A. and Bildsten, Lars and Cantiello, Matteo and Dotter, Aaron and Farmer, R. and Goldberg, Jared A. and Jermyn, Adam S. and Kanbur, S. M. and Marchant, Pablo and Thoul, Anne and Townsend, Richard H. D. and Wolf, William M. and Zhang, Michael and Timmes, F. X.},
title = {Modules for Experiments in Stellar Astrophysics (MESA): Pulsating Variable Stars, Rotation, Convective Boundaries, and Energy Conservation},
journal = {The Astrophysical Journal Supplement Series}
}

@ARTICLE{Kovetz1994,
       author = {{Kovetz}, Attay and {Prialnik}, Dina},
        title = "{Accretion onto a 1.4 M$_{sun}$ White Dwarf: Classical Nova, Recurrent Nova, or Supernova?}",
      journal = {\apj},
         year = 1994,
        month = mar,
       volume = {424},
        pages = {319},
          doi = {10.1086/173891},
       adsurl = {https://ui.adsabs.harvard.edu/abs/1994ApJ...424..319K}
}

@ARTICLE{Hachisu1999,
       author = {{Hachisu}, Izumi and {Kato}, Mariko and {Nomoto}, Ken'ichi and {Umeda}, Hideyuku},
        title = "{A New Evolutionary Path to Type IA Supernovae: A Helium-rich Supersoft X-Ray Source Channel}",
      journal = {\apj},
         year = 1999,
        month = jul,
       volume = {519},
       number = {1},
        pages = {314-323},
          doi = {10.1086/307370},
archivePrefix = {arXiv},
       eprint = {astro-ph/9902303},
 primaryClass = {astro-ph},
       adsurl = {https://ui.adsabs.harvard.edu/abs/1999ApJ...519..314H}
}

@ARTICLE{Han2004,
       author = {{Han}, Z. and {Podsiadlowski}, Ph.},
        title = "{The single-degenerate channel for the progenitors of Type Ia supernovae}",
      journal = {\mnras},
         year = 2004,
        month = jun,
       volume = {350},
       number = {4},
        pages = {1301-1309},
          doi = {10.1111/j.1365-2966.2004.07713.x},
archivePrefix = {arXiv},
       eprint = {astro-ph/0309618},
 primaryClass = {astro-ph},
       adsurl = {https://ui.adsabs.harvard.edu/abs/2004MNRAS.350.1301H}
}

@ARTICLE{Kato2004,
       author = {{Kato}, Mariko and {Hachisu}, Izumi},
        title = "{Mass Accumulation Efficiency in Helium Shell Flashes for Various White Dwarf Masses}",
      journal = {\apjl},
         year = 2004,
        month = oct,
       volume = {613},
       number = {2},
        pages = {L129-L132},
          doi = {10.1086/425249},
archivePrefix = {arXiv},
       eprint = {astro-ph/0407632},
 primaryClass = {astro-ph},
       adsurl = {https://ui.adsabs.harvard.edu/abs/2004ApJ...613L.129K}
}

@ARTICLE{Chen2022,
       author = {{Chen}, Hai-Liang and {Tauris}, Thomas M. and {Chen}, Xuefei and {Han}, Zhanwen},
        title = "{Formation of the Double White Dwarf Binary PTF J0533+0209 through Stable Mass Transfer?}",
      journal = {\apj},
         year = 2022,
        month = jan,
       volume = {925},
       number = {1},
          eid = {89},
        pages = {89},
          doi = {10.3847/1538-4357/ac3bb6},
       adsurl = {https://ui.adsabs.harvard.edu/abs/2022ApJ...925...89C}
}

@misc{GaiaDR3-2022,
  author = {{Gaia Collaboration}},
  year = {2022},
  title = {Gaia Data Release 3 (Gaia DR3) Part 1 Main source},
  howpublished = {VizieR Online Data Catalog, I/355},
  url = {https://vizier.cds.unistra.fr/viz-bin/VizieR-3?-source=I/355},
}
\bibliographystyle{aasjournal}

\end{document}